\documentclass{article}
\usepackage{graphicx} 
\usepackage{microtype} 
\usepackage{changepage} 
\usepackage{enumitem}   
\usepackage{amsmath}    
\usepackage{natbib}
\usepackage{url}
\usepackage{authblk}
\usepackage{hyperref}
\usepackage{amsmath}

\title{Co-Developing Causal Graphs with Domain Experts Guided by Weighted FDR-Adjusted p-values}
\date{September 2024}
\author [1]{Eli Y. Kling
\thanks{Email: eli.kling@avanade.com}
\thanks{Alternate Email: eli\_kling@hotmail.com}
\thanks{LinkedIn: \href{https://www.linkedin.com/in/elikling/}{https://www.linkedin.com/in/elikling/}}}
\affil[1]{Avanade, London, UK}

\newenvironment{indentpara}
    {\begin{adjustwidth}{2em}{}}
    {\end{adjustwidth}}

\begin{document}

\maketitle

\begin{abstract}
This paper proposes an approach facilitating co-design of causal graphs between subject matter experts and statistical modellers. Modern causal analysis starting with formulation of causal graphs provides benefits for robust analysis and well-grounded decision support. Moreover, this process can enrich the discovery and planning phase of data science projects.

The key premise is that plotting relevant statistical information on a causal graph structure can facilitate an intuitive discussion between domain experts and modellers. Furthermore, Hand-crafting causality graphs, integrating human expertise with robust statistical methodology, enables ensuring responsible AI practices. 

The paper focuses on using multiplicity-adjusted p-values,  controlling for the false discovery rate (FDR), as an aid for co-designing the graph. A family of hypotheses relevant to causal graph construction is identified, including assessing correlation strengths, directions of causal effects, and how well an estimated structural causal model induces the observed covariance structure.

An iterative flow is described where an initial causal graph is drafted based on expert beliefs about likely causal relationships. The subject matter expert's beliefs, communicated as ranked scores could be incorporated into the control of the measure proposed by Benjamini and Kling, the FDCR (False Discovery Cost Rate). The FDCR-adjusted p-values then provide feedback on which parts of the graph are supported or contradicted by the data. This co-design process continues, adding, removing, or revising arcs in the graph, until the expert and modeller converge on a satisfactory causal structure grounded in both domain knowledge and data evidence.

\end{abstract}

\section{Introduction}

The current common practice for conducting data science development projects is to kick off with a "Discovery" phase, where the problem is defined. This phase is often conducted in a workshop setting, where the data scientist and the subject matter expert work together to specify the target, postulate potential drivers or features and identify the data sources. Many frameworks specify a 'hypothesis workshop' as part of the kick-off and shaping steps. These sessions are usually conducted before the statistical modeller sees the data and are typically meant as a vehicle for identifying and prioritising relevant data sources. This practice encourages blowing up the list of features to "through" at a model. Automatic feature selection and modelling tools have been developed to address the tsunami of features.

The increasing reliance on complex models and automated data-mining systems raises important questions about interpretability, inference and correct methodology implementation [\cite{pearlMackenzie2018}]. For example, \cite{andersonCook2001} voices a paradigm common to statistical modelling and data mining: perform an automatic variable selection and then allow the expert to overlay the business or physical context. This practice results in models that are not easily explained. Thus also posing a challenge to implementing responsible AI guidelines.

The issues and concerns above are exacerbated when the models are used to drive decisions and actions. The nature of many challenges in the business and manufacturing worlds is such that it is often impossible to design experiments to cleanly assess the impact of decisions and actions on outcomes. For instance, the Market Mix Methodology (MMM) attempts to attribute success to marketing activities in a world of confounding co-activities and poor data [\cite{woodall2000}]. In cases where experimentation is not possible, there is a risk that misspecification and misuse of confounders will not only result in wrong action-to-outcome impact estimates but could also indicate the wrong direction of effect, as demonstrated by Simpson's paradox. As \cite{peters2017} state, ``\textit{The Simpson's paradox is not so much of a paradox but rather a warning of how sensitive causal reasoning can be with respect to model misspecifications.}'' A robust approach to address these situations is to deploy techniques developed in the field of statistical causality analysis [\cite{pearlMackenzie2018}].

Modern statistical causal analysis, exemplified by Pearl and Mackenzie \cite{pearlMackenzie2018}, begins with the formulation of a causal graph. This approach offers several benefits to the quality and robustness of analysis and decision support. The causal graph serves as an intuitive tool that bridges the knowledge and beliefs of a subject matter expert (SME) with the statistical modeller's data-driven insights. More importantly, Pearl \cite{pearl2014} emphasises that mapping causal mediating factors illuminates 'intrinsic properties of reality that have tangible policy implications'.

\cite{shmueli2010} discusses the tension between the two analysis goals explainability and productiveness.Each has a different starting point and nuances of the methodology used. Specifying correctly a causal graph has the benefit of expalianbility and bringing to the for potential introduction of bias that is not compliant with responsible AI guidelines. Once an explainable model is crafted the statistical modeller may proceed to fit a prediction oriented model guided by the insights of the discovery step. This is in contrast to \cite{breiman2001} "Using complex predictors may be unpleasant, but the soundest path is to go for predictive accuracy first, then try to understand why"

A growing body of research explores graphical methods to visualise and communicate the results of causality analysis. For instance, \cite{hoque2022} discuss supporting lay users with no specific expertise in machine learning, promoting an interactive approach aimed at "emotionally connecting" the subject matter expert.

While the field of automatically discovering causal graphs is active [\cite{kaiser2021}], this paper focuses on an approach to aid discussion between the statistical modeler and the subject matter expert while constructing a causality graph. The discussion explores the formation of a causality graph as a tool to be used during the discovery and exploratory data analysis phase.

\begin{figure}[h]
\centering
\includegraphics[width=0.8\textwidth]{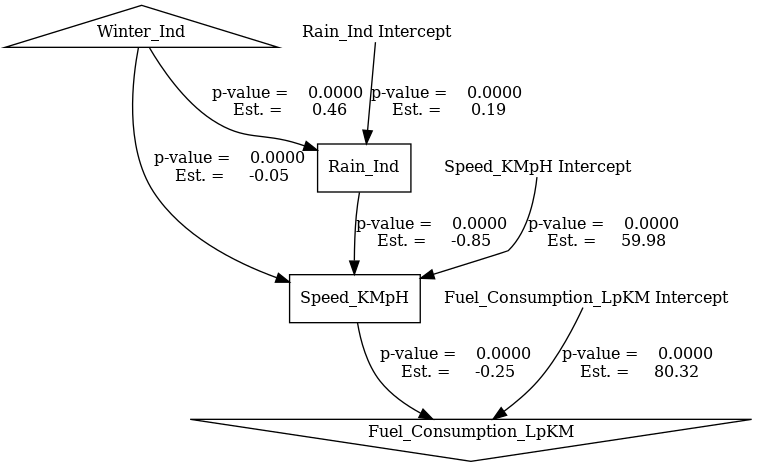}
\caption{An example of presenting a causality graph overlaid with parameter estimates and FDR-corrected p-values}
\label{fig:causality_graph_example}
\end{figure}

This paper posits that plotting relevant information on a causality graph facilitates discussion between the statistical modeller and the subject matter expert. This approach does not argue for setting aside rigour for the sake of simplifying the discussion. The False Discovery Rate (FDR) or False Discovery Cost Rate (FDCR) is an appropriate measure for introducing multiplicity considerations in an intuitive way. Figure \ref{fig:causality_graph_example} demonstrates that this type of presentation is self-explanatory and suitable for both systematic-level discussions with subject matter experts and modelling-level discussions with trained statisticians.

This paper proposes an approach to guide the construction of a causality graph that will underpin inference, estimation, and predictions. For simplicity but without loss of generality, the approach describes stepwise crafting of a Structural Equation Model (SEM) or Structural Causal Model (SCM) where the adjustment of the False Discovery Rate (FDR) is deployed as an aid and guide [\cite{benjaminiHochberg1995}].

The paper starts with discussing which hypotheses are relevant to the co-design of a causality graph (Section \ref{sec:relevant_hypotheses}). Once a family of hypotheses is defined, Section \ref{sec:fdr_control} describes the control of the False Discovery Rate (FDR) and its generalisation to FDCR using costs of false alarms. Section \ref{sec:toy_example} demonstrates the use of such adjustment in the construction of a Structural Causal Model (SCM) using a toy example. The paper concludes with Section \ref{sec:conclusion}, touching on thoughts not explored here.

\section{Identifying the Family of Hypotheses}
\label{sec:relevant_hypotheses}

This paper follows a paradigm of collaboratively working with subject matter experts (SMEs) to translate their expertise and experience into a hypothesised causality graph. It is natural to transform this process into a set of hypotheses that reflect the SME's beliefs. The next logical step is applying some multiplicity control. The False Discovery Rate (FDR) is a natural multiplicity measure for this setting.

Before delving into multiplicity error measures and their control, it is important to define the family of hypotheses and understand the co-dependence between the hypotheses, the statistics, and the resulting p-values. [\cite{Bromberg2009}] discuss using Pearl's theorems on the properties of conditional independence relations [\cite{Pearl1988}] and avoiding the execution of some statistical tests to reduce the computational load. A note of caution: avoiding calculating the test does not exclude the hypotheses from the family if it is pertinent to structuring the causality graph unless the correlation to another hypothesis to another hypothesis is 1.

In the context of assessing of a causality graph, hypotheses are postulated based on SME's experience, their prior beliefs, and suggestions derived from explanatory analysis. As \cite{Pena2008} points out, the hypotheses focus on the discovery of edges in the Directed Acyclic Graph (DAG). 

The identification of the relevant hypothesis is somewhat derived from the typical co-design process. Note how the choice of modelling is incorporated here:

\begin{enumerate}
    \item Identify the outcomes of interest and decisions that are likely to affect the outcomes
    \item Use the Six Sigma fish--bone diagram to list potential drivers of the outcomes
    \item Obtain, clean, and prepare data for the information listed in the fish--bone diagram
    \item Calculate the correlations and their corresponding p-values (The null hypothesis for each pair is that there is no correlation or co-dependence)
    \item Assign a causal direction to each pair (could also be 'no causal relationship')
    \item Assign a belief score to each postulated causal relationship. For instance:
    \begin{indentpara}
        0 = Causal relation is not possible
        
        1 = Could be a causal relationship
        
        2 = It is likely this is a causal relationship
        
        3 = Known causal relationship
        
    \end{indentpara}
    
    \item Using the causality graph defined by (4) \& (5), fit a Structural Causal Model (SCM). This naturally defines a sub-family of hypotheses:
    \begin{enumerate}
        \item The model does not explain the data (tested with a set of fit statistics)
        \item The model coefficients are zero
        \item The mean of the residuals is zero (could also hypothesise on the variance of the residuals)
    \end{enumerate}
    \item A further set of hypotheses are based on the assertion that the correlations induced by the SCM in (7) are the same as seen in (4). This is in line with the field of structure learning. For example, \cite{Gasse2015} advise that a DAG should be as close as possible to the global dependence structure
\end{enumerate}

Step (8) is linked notionally to the definition of faithfulness. A graph is faithful to some distribution if the graph connectivity represents exactly the dependencies and independences dictated by the distribution (see for instance, \cite{Bromberg2009}.)

Moreover, rather than examining the correlations, it makes sense to work with covariances so that the direction is also considered.

The hypotheses in (8) are an important feedback to the co-design process as 'unattended' correlations would be highlighted. However, the way (8) is formulated above results in counter-intuitive p-values (big is good) and does not really prove the SCM is reflecting the covariances and variances. Rather, it only shows that there is no evidence to refute that it does. (8) should be couched similarly to equivalence testing, where the null hypotheses are that the covariances and variances induced by the SCM are different from the observed correlations:

\begin{equation}
H_0^i: |\rho_{scm}^i - \rho_{data}^i| \geq \delta
\end{equation}
\begin{equation}
H_1^i: |\rho_{scm}^i - \rho_{data}^i| < \delta
\end{equation}

Where, for $n$ variables, $\rho_{data}^i$ ($i = 1, \ldots, \frac{n(n+1)}{2}$) are the observed correlations in the data. $\rho_{scm}^i$ ($i = 1, \ldots, \frac{n(n+1)}{2}$) are the matching correlations induced by the SCM and $\delta$ is a threshold parameter. It is customary to form the above as two one-sided hypothesis tests with two p-values that should be small if the two measures are similar.

Examining the p-values in (7) and (8) informs suggested changes to the causality graph. Steps (7) and (8) are repeated until the Modeller and the SME are satisfied. Thus, multiplicity of hypothesis control is called for. As the process could be viewed as a screening process where the proportion of true null hypotheses is small, the control of the False Discovery Rate (FDR) is appropriate. I.e., the p-values reviewed during the process should be FDR adjusted. We propose that due to the consistency property of the FDR adjustment, it is sufficient to consider the adjustment for each iteration separately. Moreover, the FDR adjustment provides a mechanism for inclusion of the belief scores where they can define the weights for a weighted FDR adjustment.

Another source of hypotheses comes from the field of causality structure learning where conditional independence tests underpin a stepwise (forward or backward) causality graph construction. This paper is concerned with "manual" co-design of the causality graph rather than an automated, data-driven algorithm. Thus, the hypotheses in (8) are sufficient for driving and informing the discussion between the SME and the Statistical Modeller, although not as nuanced as the conditional independence tests.

It is important to understand the correlation structure among all the p-values used, as some of the procedures for the control of the FDR (e.g. the Bejamini-Hochberg procedure [\cite{benjaminiHochberg1995}]), control the FDR only under independence or under Positive Regression Dependence (PRDS). In cases where complex correlations are suspected, it is possible to use the Benjamini-Yekutieli algorithm [\cite{benjaminiYekutieli2001}]  or bootstrap as demonstrated by \cite{yekutieli1999} using the foundations laid by \cite{westfallyoung1992}.

A nuance of the above requirement for independence or at least PRDS is that it is sufficient to show it for the true null hypothesis. As the hypotheses pertain to the existence of a causal effect manifested by conditional dependence, the true null hypotheses that could be interdependent are those considering the same edge. Thus, there might be a correlation for the p-values assessing the coefficient for modelling an edge and the p-value for the correlation induced by the SCM for that edge. Arguably, they are measuring overlapping constructs and thus are PRDS. However, the hypotheses in (8) are formulated as an equivalence between the correlation structure induced by the SCM and the observed correlation. For a world where \textit{A} does not directly cause \textit{B}, the true causality graph will not have an edge linking nodes \textit{A} and \textit{B} (edge \textit{AB}). For an SCM containing an edge \textit{AB}, the hypothesis that the coefficient for the regression of \textit{B} on \textit{A} is \textit{zero} is true. However, the hypothesis that the induced correlation between \textit{A} and \textit{B} is equivalent to the observed could be true or false as the correlation may be induced. Simple cases where a null coefficient is fully informative of the existence of the correlation can be constructed. Arguably these are simple cases where a model crafting session is not required. Therefore, the p-values considered here could be treated as independent or PRDS. 

For illustration purpose, consider a simple causality Graph with one mediator T $\rightarrow$ M; T $\rightarrow$ O; M $\rightarrow$ O. Assuming T, M \& O are continuous and the functional form linking them is linear, the Structural Equation Model (SEM) could be:

\begin{equation}
\hat{M} = \beta_0 + \beta_1 T + \varepsilon_1
\end{equation}
\begin{equation}
\hat{Y} = \beta_2 + \beta_3 T + \beta_4 M + \varepsilon_2
\end{equation}

Where $\varepsilon_j \sim N(0, \sigma_j)$; $j = 1, 2$

The hypotheses of interest are:

\begin{enumerate}[label=(\roman*)]
    \item $H_{0,i}^{Parms}: \beta_i = 0$;
    
    $H_{1,i}^{Parms}: \beta_i \neq 0$;
    
    $\forall i \in \{0, 1, 2, 3, 4\}$
    \item $H_{0,j}^{Noise}: \varepsilon_j \sim N(0, .)$;
    
    $H_{1,j}^{Noise}: \varepsilon_j \not\sim N(0, .)$;
    
    $\forall j \in \{1, 2\}$
    
    \item $H_{0,x,y}^{Cov}: |\text{SEM induced Cov(x,y)} - \text{observed  Cov(x,y)}| > \delta$; 
    
    $H_{1,x,y}^{Cov}: |SEM \text{ induced } Cov(x,y) - \text{observed } Cov(x,y)| \leq \delta$; 
    
    $\forall x, y \in \{T, M, Y\}$ – upper triangle \& diagonal
    
\end{enumerate}

(i) asserts that the coefficients are not zero. For simple linear modelling, the p-values could be theoretically derived; (ii) tests the assumption that the noise is Gaussian distributed. This could be tested using a Chi-square test. It could be argued that the roles of $H_0$ and $H_1$ should be flipped, posing a challenge of testing for "non-normality".

The process we describe is iterative. At each step, edges could be added or removed and the coefficients refitted (either backwards or forwards construction). It could be argued that the hypotheses of all the steps should be considered as one family. The consistency property of the FDR (when Number of True null hypotheses: $n_0 \ll$ Number of false null hypotheses: $n_1$) allows for working within iteration. Thus circumventing confusion due to having several p-values generated for the same hypothesis. Therefore, It is sufficient to control the FDR within each iteration.

Before providing an example for the process, the FDR and the False Discovery Control Rate (FDCR) and their control are discussed in the next section.

\section{False Discovery [Cost] Rate}
\label{sec:fdr_control}

This section is based on technical papers by \cite{benjaminiKling2005} and \cite{kling2005phd}.

When testing several hypotheses simultaneously to reach an overall decision, there is a trade-off between controlling the type I error for per-hypothesis considerations and the overall hypothesis, which is usually their union. This issue of balancing error rates is known as the Multiplicity Problem. This dilemma is encountered almost universally where statistics are applied. Some of the most prominent statisticians have addressed it. However, there is neither a common nor global approach. For instance, \cite{tukey1994} discussed a special case of the problem, pairwise comparisons. He continued discussing it as late as 1991 [\cite{Tukey1991}]. An extensive overview of the research of the Multiplicity problem may be found in \cite{Hochberg1987}, \cite{westfallyoung1992}, \cite{benjaminiHochberg1995}, \cite{Hsu1996}, and \cite{Westfall1999}.

\cite{westfallyoung1992} demonstrated this: "For example, a particular survey may identify a small p-value, say p = .005, and claim that the associated effect is 'statistically significant.' This p-value is interpretable as follows: when there is no causal basis for the effect, there is only a 0.5\% probability of observing a result as extreme as the observed result. On the other hand, it is possible that the multiplicity adjusted p-value is .15 (Adj-p = .15), which is not statistically significant. This adjusted p-value incorporates the multiple tests performed and can be interpreted as follows: when there is no causal basis for any effect tested, there is a 15\% chance that somewhere in the experiment a result as extreme as the observed result of .005 will appear."

Classical procedures aim to control the probability of committing at least one type-I error when considering a family of hypotheses simultaneously to control the multiplicity effect. The control of this Familywise Error Rate (FWE, see Equations \ref{eq:strong_fwe} \& \ref{eq:weak_fwe}) is usually required in the strong sense (Equation \ref{eq:strong_fwe}), i.e., under all configurations of true and false hypotheses [\cite{Hochberg1987}]. The main problem with classical methods is that they tend to have low power. Consequently, it has been argued that no special control is needed (e.g. [\cite{Rothman1990, Saville1990}]).

An alternative, more powerful measure was introduced by \cite{benjaminiHochberg1995}: the False Discovery Rate (FDR, Equation \ref{eq:fdr}). It is an appropriate error rate to control in many problems where the (strong) control of the FWE is not needed. The FDR is the expected ratio of the number of erroneous rejections to the number of rejections (discoveries) and is equal to or less than the FWE. The two error rates are equal when the number of true null hypotheses ($m_0$) equals the number of hypotheses under test ($m$). When $m_0 < m$, the FDR may be substantially lower than the Familywise Error Rate, so an FDR controlling procedure at conventional levels can be more powerful. \cite{benjaminiHochberg1995} provided a linear step-up procedure (BH) that controls the FDR for independent test statistics. \cite{benjaminiYekutieli2001} showed that when the test statistics are PRDS correlated, the BH procedure controls the FDR. Furthermore, they introduced a resampling-based procedure that controls the FDR [\cite{yekutieli1999}].

A useful aspect of the BH procedure is that it provides quite consistent discoveries when applied to subsets of the family of hypotheses. For example, when evaluating individual hypotheses pertaining to enumeration districts, the family could be the whole of the United States or just a specific state. FWE control procedures usually will provide conflicting individual decisions for the different hypotheses families, whereas the FDR control will generate consistent discoveries when $m_0 \ll m$.

\cite{Efron2001}, \cite{Efron2002}, \cite{Storey2002}, and \cite{Tang2005, Tang2007} frame the control of the FDR using a Bayesian paradigm. This approach allows the discussion of a prior belief on the probability that the null hypothesis is correct. This is somewhat related to the discussion in this paper of assisting experts in mapping out their beliefs on causality.

Following the notation used by \cite{benjaminiHochberg1995, benjaminiHochberg1997} and \cite{benjaminiKling2005}: For a composition of $m$ sub-hypotheses $(H_{0i}; i=1,2,\ldots,m)$ let the intersection hypothesis be $H_{00} = \bigcap_{i=1}^m H_{0i}$. Let $R_i$ $(i=0,1,2,\ldots,m)$ be 1 if $H_{0i}$ is rejected and zero otherwise; and let $V_i$ $(i=0,1,2,\ldots,m)$ be 1 if $H_{0i}$ is erroneously rejected and zero otherwise. Note, if $H_{0i}$ is true then $V_i=R_i$. Furthermore, let the number of rejections ("Discoveries") be $R = \sum_{i=1}^m R_i$, and the number of erroneous rejections be $V = \sum_{i=1}^m V_i$. Note that $R$ and $V$ do not include $R_0$ and $V_0$. Let $I_R=1$ when $R>0$ otherwise $I_R=0$. Similarly, $I_V=1$ when $V>0$ otherwise $I_V=0$. To complete the notation used in this paper, define $I_0$ as the set of indices of the true null sub-hypotheses $(I_0=\{j;H_{0j} \text{ is true}, 1 \leq j \leq m\})$; and the number of true null sub-hypotheses is $m_0=||I_0||$.

The family error measures could be defined using the above terms:

\begin{align}
\text{Strong-FWE} &= P(V>0) = E[\max V_j; 1\leq j\leq m] \label{eq:strong_fwe} \\
\text{Weak-FWE} &= E[V_0] = E[V_0/R_0] \label{eq:weak_fwe} \\
\text{FDR} &= E[V/R] \label{eq:fdr} \\
\text{Weighted FDR} &= \text{WFDR} = E\left[\frac{\sum_{i=1}^m w_i V_i}{\sum_{i=1}^m w_i R_i}\right] \label{eq:wfdr}
\end{align}

where $V/R$ and $V_0/R_0$ are defined as zero when $V=R=0$ and $V_0=R_0=0$ respectively.

The FWE is appropriate when even one erroneous discovery is not desired. Procedures such as the Bonferroni procedure, Holm's procedure, Hochberg's Procedure [\cite{hochberg1988}], and Tukey's  T-method for pairwise comparisons [\cite{tukey1994}], all control the FWE in the strong sense (Equation \ref{eq:strong_fwe}). This type of control is relevant to situations where any erroneous discovery implies a very high cost. For example, such conservativeness is required when examining the primary end-points during Phase III clinical trials.

Control of the FWE in the weak sense (Equation \ref{eq:weak_fwe}) is achieved by testing directly the intersection null hypothesis (and not controlling the individual hypothesis). For instance, by using the multivariate Hotelling $T^2$ statistic. Thus, the overall type I error rate is controlled only when all the sub-hypotheses are true $(m_0 = m)$. This situation is very common in Statistical Process Control (SPC); where once an out-of-control signal is given (the intersection hypothesis is rejected) it is assumed that it is no longer necessary to protect from erroneous sub-discoveries, and on the other hand increased power is desired. 

Generally, the use of the FDR (Equations \ref{eq:fdr} \& \ref{eq:wfdr}) is appropriate in situations where high power is imperative and a pre-specified percent of the wrongly rejected (individual) hypotheses is tolerable and does not affect the quality of the overall decision. This is usually characterised by the belief that $m_0 \ll m$. The FDR may be used in pilot or screening studies, grouping analysis, and situations where many variables are considered such as data-mining and Bioinformatics. In such situations, the error from a single erroneous rejection is not always as crucial for drawing conclusions for the family hypothesis. Thus, we are ready to bear more errors when many hypotheses are rejected, but with less when fewer are rejected. The last notion is reflected by the control of the FDR - for which one must specify the acceptable expected proportion of wrong discoveries. For an example, see \cite{Grigg2008} for a review of papers that proposed adopting the FDR for multiple CUSUM charts.

Though widely researched, there are no clear-cut rules for which error measure to use and what multiplicity control procedure to use. \cite{benjaminiKling2005} and \cite{kling2005phd} argued that this decision is part of the modelling process. They show that in situations where costs or weights can be attributed to erroneous discoveries, the weak-FWE and the FDR are special cases of a generic cost-based error measure, the False Discovery Cost Ratio (FDCR, Equations \ref{eq:fdcr}). They show that the W-BH procedure [\cite{benjaminiHochberg1997}] keeps its control also when the test statistics are PRDS ON $I_0$ and propose a procedure for the control of the FDCR when the test statistics are PRDS.

Assigning variable cost of $C_i$ $(i=0,1,2,\ldots,m)$ of an individual erroneous discovery (e.g., rejecting $H_{0i}$ results in stopping machine $i$ for maintenance) and a fixed cost $C_0$ for the overall discovery (rejecting $H_{00}$ results in calling in an engineer), the cost of false discoveries is $C_0 I_{V_0} + \sum_{i=1}^m C_i V_i$. In the spirit of the FDR and using the above notation, the expected proportion of the cost of false discoveries is:

\begin{equation}
E\left[\frac{C_0 I_{V_0} + \sum_{i=1}^m C_i V_i}{C_0 I_{R_0} + \sum_{i=1}^m C_i R_i}\right] \leq  E\left[\frac{C_0 I_V + \sum_{i=1}^m C_i V_i}{C_0 I_R + \sum_{i=1}^m C_i R_i}\right], \label{eq:fdcr_proportion}
\end{equation}

where the proportion is zero when the denominator is zero and $I_R=I_{\{R>0\}}$ and $I_V =I_{\{V>0\}}$. 

\cite{benjaminiKling2005} define the False Discovery Cost Rate as the expected ratio of the cost wasted due to erroneous discoveries to the total cost related to the discoveries:

\begin{equation}
\text{FDCR} = E\left[\frac{C_0 I_{V_0} + \sum_{i=1}^m C_i V_i}{C_0 I_{R_0} + \sum_{i=1}^m C_i R_i}\right] = E\left[\frac{\sum_{i=0}^m C_i V_i}{\sum_{i=0}^m C_i R_i}\right], \label{eq:fdcr}
\end{equation}
where the proportion is to be defined as zero when the denominator is zero.

The Weak FWE, the FDR, and the W-FDR are special cases of this measure obtained by specific structures of the costs. When $C_0=0$ the FDCR is the W-FDR. The FDR is obtained by further assigning equal costs to the $m$ hypotheses. On the other hand, when the cost consists only of $C_0$ the FDCR is the FWE in the weak sense. 

It is interesting to examine the meaning of controlling the FWE in the strong sense from the viewpoint of the FDCR. Seeking to control the probability of making any false rejection suggests that every erroneous discovery is very costly and perceived as infinite. Thus, the Strong-FWE can be approximated by the FDCR, but it cannot be expressed in the context of additive costs. The Strong-FWE may be expressed in terms of "relative cost", $E\left[\frac{max(v_i)}{max(R_i}\right]$, which does not render itself easily to economic interpretation.

\cite{benjaminiKling2005} propose testing simultaneously $H_{00}$ and the rest of the sub-hypotheses, using the FDCR to correct for multiplicity. In particular, testing $H_{00}$ through Hotelling's $T^2$, weighted Simes' statistic (Equation \ref{eq:weighted_simes}), and Fisher's statistic (Equation \ref{eq:fisher})

\begin{equation}
P^{ws} = \min_j \left[\frac{\sum_{i=1}^m c_i}{\sum_{i=1}^j C_{(i)}} P_{(j)}\right]. \label{eq:weighted_simes}
\end{equation}

\begin{equation}
P^F = -2\sum_{i=1}^m \ln p_i. \label{eq:fisher}
\end{equation}

They showed that an FDCR controlling procedure could be constructed by applying the W-BH procedure to the $m+1$ p-values $P_0, P_1, \ldots, P_m$, where the weighted Simes adjusted p-value for $P_0$ (corresponding to $H_{00}$) is used. Thus, the control of the FDCR could be achieved when the p-values associated with the $m_0$ true hypotheses are independent or PRDS.

\section{Constructing a Causality Graph while controlling the FDR}
\label{sec:toy_example}

The process of constructing a causality graph and a statistical model should involve a statistically trained data scientist and a subject matter expert (SME) working through a process similar in spirit to variable selection. They identify the target variables, potential explanatory variables, and assign beliefs to the direction of causality. These beliefs could be expressed as a score. Then the statistician prepares the data for analysis and iteratively constructs the causality graph and models with the SME.

\cite{foygelBarber2015} propose an approach for variable selection in linear models that controls the False Discovery Rate (FDR) where the discoveries are the selected variables to be included in the model. They point out that in many practical situations, there are only a few relevant variables among the many recorded. It makes sense to view the process of constructing the causality graph as a systematic screening of variable selection hypotheses pertaining to strengths of relationships and direct and indirect causality effects.

To illustrate such a process, a synthetic example was generated. The demonstration will iteratively create a Structural Causality Model where feedback to the SME is a graphical representation overlaid with the False Discovery Cost Rate (FDCR) adjusted p-values. The toy example describes an offshore wind farm operator who decides how fast to allow the turbine to turn. The key causality relationships of interest are "Rotational\_RPM" $\rightarrow$ "Energy\_Yield" and "Rotational\_RPM" $\rightarrow$ "Perceived\_Noise". Appendix A describes the construction of the toy example. Table \ref{tab:variables} lists the variables the SME suggested could have bearing on the outcome. 20,000 observations were generated.

\begin{table}[htbp]
\centering
\caption{Variables that might have bearing on the 'Financial Outcome'}
\label{tab:variables}
\begin{tabular}{p{3.5cm}p{7.5cm}}
\hline
Variable & Description \\
\hline
Winter Ind & 1 if winter season, 0 otherwise \\
Sea Temperature & Water surface temperature \\
Wind Speed & Wind speed measured at a specified height \& location \\
Strength Degradation & Mechanical degradation of the bearings\\
Rotational RPM & Action variable – an operational decision \\
Energy Yield & Output to maximize \\
Perceived Noise & Outcome of interest to keep under a threshold. Calculated as per relevant ISO and regulations \\
\hline
\end{tabular}
\end{table}

The univariate exploratory analysis kicks off the co-design process, facilitating conversations between the statistician and the SME. The next step is to evaluate the pairwise relationships. Table \ref{tab:correlations} lists the Pearson pairwise correlations, and Table \ref{tab:fdr_adjusted_bootstrapped_p_values} lists the FDR adjusted (BH adjusted) p-values obtained through bootstrapping the Pearson $R^2$ [\cite{benjaminiYekutieli2001}].

\begin{table}[htbp]
\centering
\caption{Pearson pairwise correlations}
\label{tab:correlations}
\scriptsize 
\begin{tabular}{p{1.5cm}p{1.5cm}p{1.5cm}p{1.5cm}p{1.5cm}p{1.5cm}p{1.5cm}p{1.5cm}}
\hline
 & Winter Ind & Sea Temperature & Wind Speed & Strength Degradation & Rotational RPM & Energy Yield & Perceived Noise \\
\hline
Winter Ind & 1.00 & -0.92 & 0.67 & -0.00 & 0.67 & 0.28 & -0.62 \\
Sea Temperature & -0.92 & 1.00 & -0.62 & 0.01 & -0.61 & -0.25 & 0.57 \\
Wind Speed & 0.67 & -0.62 & 1.00 & 0.00 & 0.99 & 0.41 & -0.93 \\
Strength Degradation & -0.00 & 0.01 & 0.00 & 1.00 & 0.00 & -0.01 & -0.00 \\
Rotational RPM & 0.67 & -0.61 & 0.99 & 0.00 & 1.00 & 0.42 & -0.91 \\
Energy Yield & 0.28 & -0.25 & 0.41 & -0.01 & 0.42 & 1.00 & -0.38 \\
Perceived Noise & -0.62 & 0.57 & -0.93 & -0.00 & -0.91 & -0.38 & 1.00 \\
\hline
\end{tabular}
\end{table}

\begin{table}[htbp]
\centering
\caption{FDR adjusted bootstrapped p-values for R2}
\label{tab:fdr_adjusted_bootstrapped_p_values}
\begin{tabular}{p{1.5cm}p{1.5cm}p{1.5cm}p{1.5cm}p{1.5cm}p{1.5cm}p{1.5cm}p{1.5cm}}
\hline
 & Sea Temperature & Wind Speed & Strength Degradation & Rotational RPM & Energy Yield & Perceived Noise \\
\hline
Winter Ind & 0 & 0 & 0.6837 & 0 & 0 & 0 \\
Sea Temperature &  & 0 & 0.5250 & 0 & 0 & 0 \\
Wind Speed &  &  & 0.8323 & 0 & 0 & 0.704 \\
Strength Degradation &  &  &  & 0.8865 & 0.575 & 0.8865 \\
Rotational RPM &  &  &  &  & 0 & 0 \\
Energy Yield &  &  &  &  &  & 0 \\
\hline
\end{tabular}
\end{table}

\begin{samepage}
Reviewing the correlations and the adjusted p-values is a good starting point for designing the first draft causality graph and attributing the SME's belief. For this example, the belief is scored simply:

\begin{indentpara}

\begin{enumerate}
\setcounter{enumi}{-1}
    \item As an SME, I do not believe there is a causal relationship between the two variables – this is implicit by not drawing an arc in the causality graph
    \item As an SME, I am not sure whether there is a causal relationship between the two variables
    \item As an SME, I believe there is a causal relationship between the two variables
    \item As an SME, I am sure there is a causal relationship between the two variables
\end{enumerate}
\end{indentpara}
\end{samepage}

The $C_i$ weights for the FDCR are set to $1/(\text{belief}+0.0001)$. All unprovided weights, such as for the tests for the intercept and $C_{00}$ (Weighted Simes), are set to 1.

Assume the SME provides the following beliefs:

\begin{verbatim}
Intercept -> Sea Temperature                  : 1
Winter Ind -> Sea Temperature                 : 3
Intercept -> Wind Speed                       : 1
Winter Ind -> Wind Speed                      : 3
Intercept -> Rotational RPM                   : 1
Strength Degradation -> Rotational RPM        : 3
Wind Speed -> Rotational RPM                  : 3
Intercept -> Energy Yield                     : 1
Rotational RPM -> Energy Yield                : 3
Wind Speed -> Energy Yield                    : 2
Sea Temperature -> Energy Yield               : 1
Intercept -> Perceived Noise                  : 1
Wind_Speed -> Perceived Noise                 : 3
Rotational_RPM -> Perceived Noise             : 3
\end{verbatim}

\begin{figure}[htbp]
\centering
\includegraphics[width=0.6\textwidth]{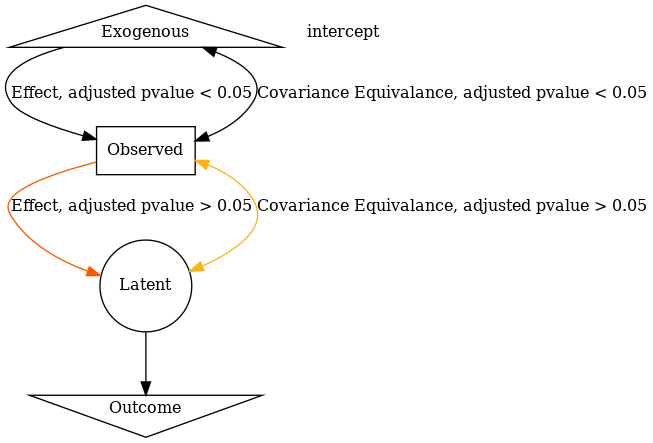}
\caption{Legend used for presenting the SCM fitting results}
\label{fig:legend}
\end{figure}

Figure \ref{fig:legend} sets the legend for the graphical representation of the results. The estimates and FDCR adjusted p-values are presented alongside the edges. The non-significant adjusted p-values are highlighted as those mark where the SME and statistician should focus their discussion.

\begin{figure}[htbp]
\centering
\includegraphics[width=1\textwidth]{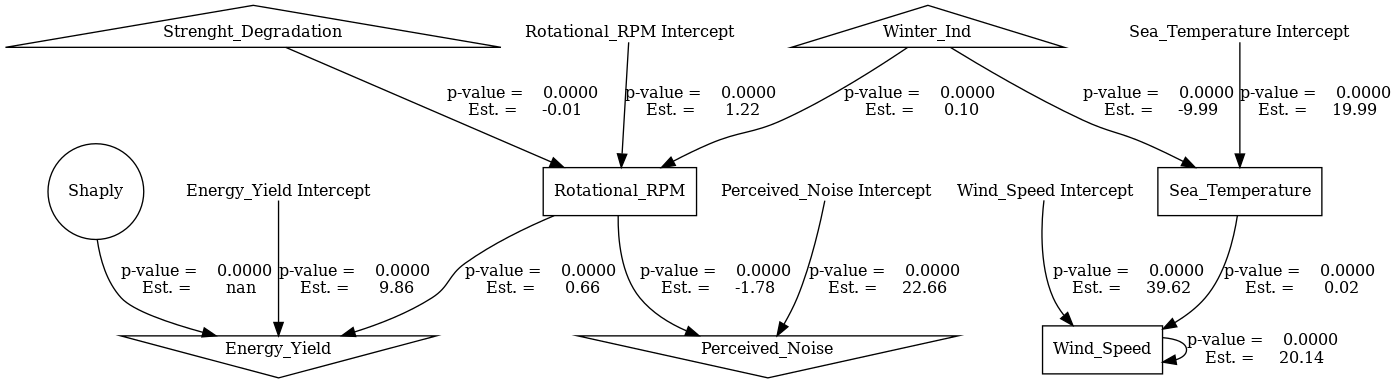}
\caption{First modeling iteration – SCM effects}
\label{fig:first_iteration}
\end{figure}

\begin{figure}[htbp]
\centering
\includegraphics[width=1\textwidth]{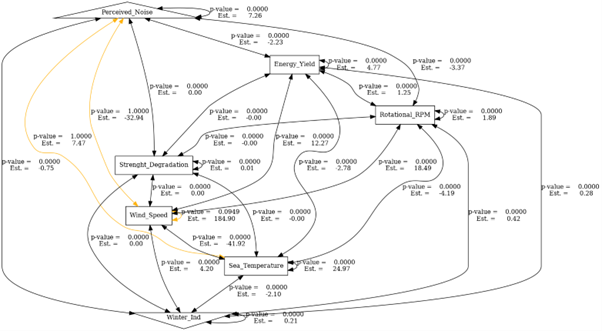}
\caption{First modeling iteration – SCM induced covariances and variances}
\label{fig:first_iteration_cov}
\end{figure}

Figure \ref{fig:first_iteration} shows a possible outcome where the significant correlations are accounted for, and the direction of the causation is informed by the SME's experience. The covariances and variances induced by the SCM model and their corresponding FDCR adjusted p-values are presented in Figure \ref{fig:first_iteration_cov}.

\begin{samepage}
The next iteration takes into account that:
\begin{itemize}
\item The estimate for Strength\_degradation $\rightarrow$ Rotational\_RPM is practically zero, albeit statistically significant.
\item The covariance between sea\_temperature and Perceived\_noise generated by the model is not equivalent to the measured relationship.
\item The covariance between wind\_speed and Perceived\_noise generated by the model is not equivalent to the measured relationship.
\end{itemize}
\end{samepage}

\begin{figure}[htbp]
\centering
\includegraphics[width=1\textwidth]{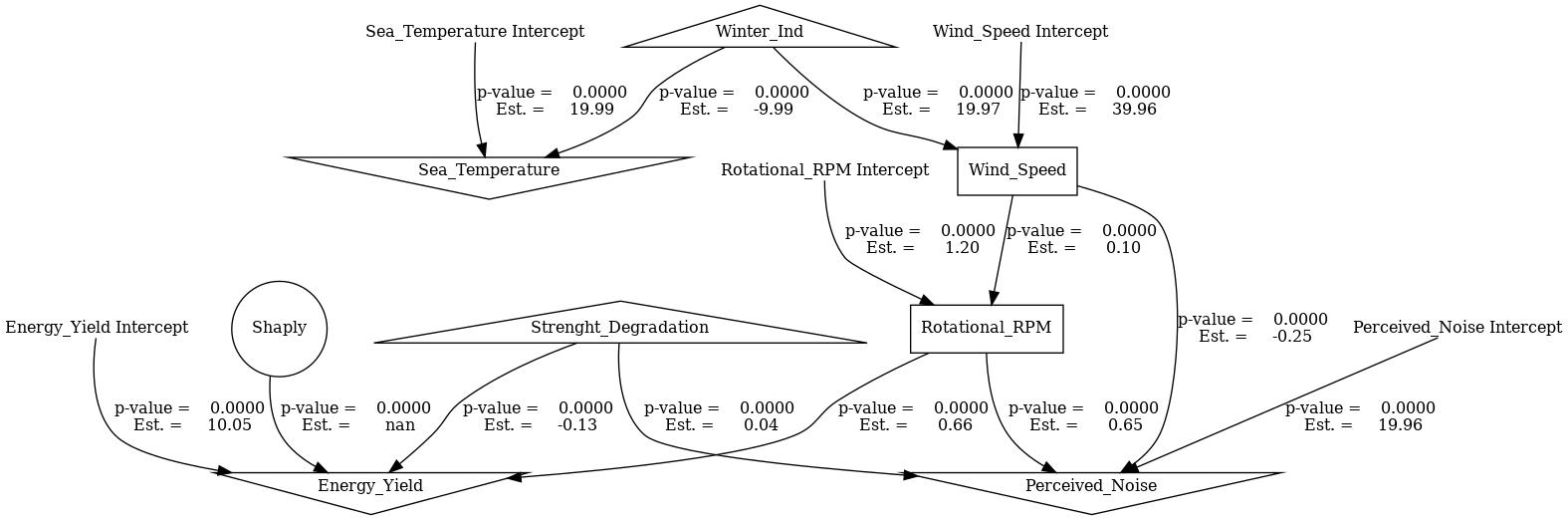}
\caption{Model results after a few design iterations – SCM effects}
\label{fig:final_model}
\end{figure}

\begin{figure}[htbp]
\centering
\includegraphics[width=1\textwidth]{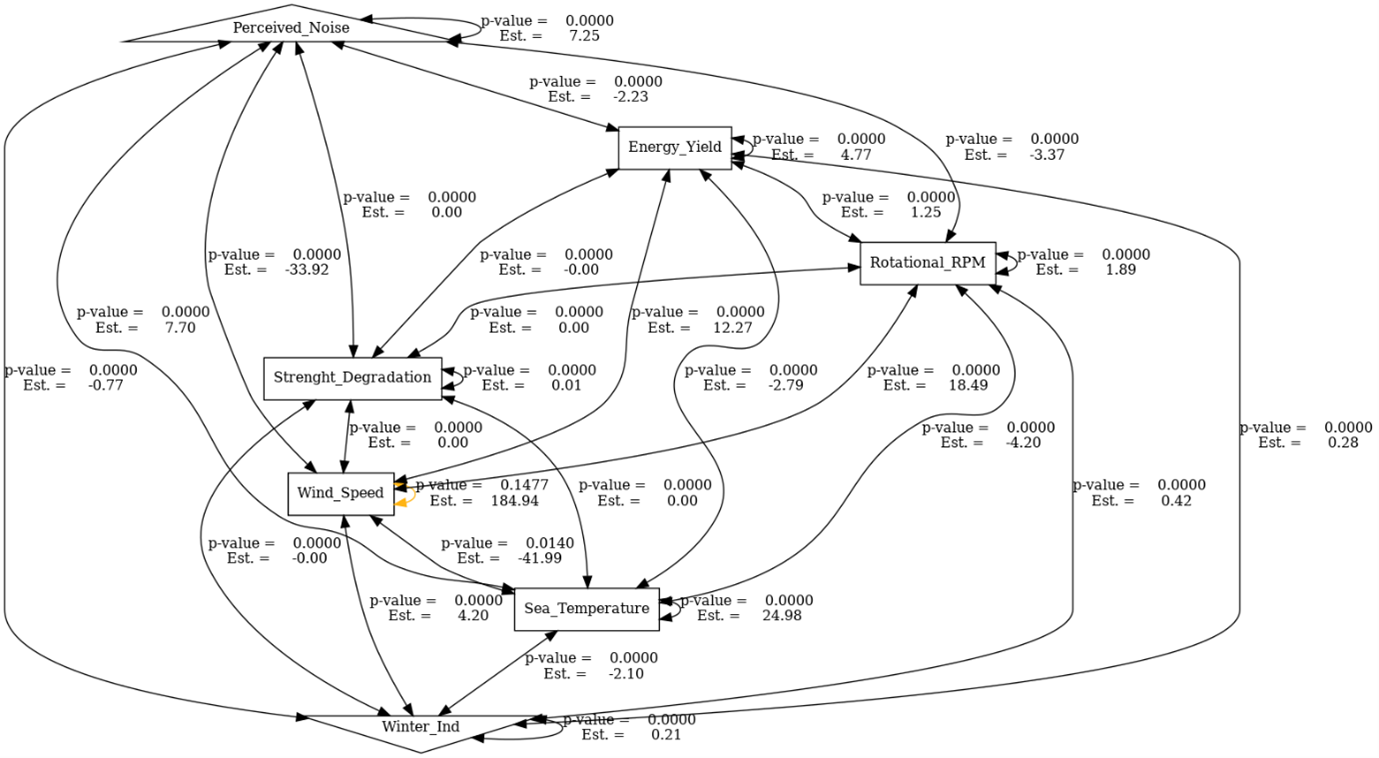}
\caption{Model results after a few design iterations – SCM induced covariances and variances}
\label{fig:final_model_cov}
\end{figure}

After several iterations, the model could look as presented in Figures \ref{fig:final_model} and \ref{fig:final_model_cov}.

\section{Conclusion}
\label{sec:conclusion}

This paper demonstrated how the FDR and FDCR could be put to use in the context of co-designing a causality graph in the discovery step of an analytical project. 

This approach favours hand crafting the causality graph but does not preclude using concepts and tools developed for automatic causality discovery. The use of FDR control during model discovery has been proposed by several researchers. For example, algorithms such as FDR-IAMB [\cite{Pena2008}] and FDR-IAPC [\cite{Gasse2014}] \& [\cite{Gasse2015}] apply techniques for controlling the multiplicity of hypotheses. These algorithms aim to learn the structure of the causality graph from the data and use FDR control as a weighting for a greedy algorithm.

The primary focus lies in facilitating communication between statistical modellers and subject matter experts (SMEs) by balancing statistical rigour with clear and intuitive presentation of results. One potential improvement would be to replace p-values with e-values as suggested by \cite{shafer2021}] and \cite{wang2022}.

While the example used employs linear regression for modelling causal relationships, the discussion is not limited to this specific functional form. Statistical modellers have the flexibility to choose the modelling approach for conditional probabilities and select appropriate measures of variable (parent node) contribution. For instance, a random forest might be used as the functional representation, with variable importance as the contribution measure.

The hypothesis testing framework described could be generalised. Instead of focusing on a specific parameter value, the null hypothesis could be formulated as: "the parent variable does not contribute to modelling the child." In this case, a nested bootstrap procedure is recommended by \cite{westfallyoung1992} and \cite{benjaminiYekutieli2001}:

1. [Outer] Bootstrap a distribution of p-values and performs FDR adjustment.

2. [Inner] Bootstrap a distribution of contribution measures or regression parameters for deriving p-values.

When using Bootstrap to obtain the distribution of the p-values. Careful thought should be given to ensuring the correlation structures under the appropriate null hypotheses is maintained.

\section*{Acknowledgements}

I would like to thank Professor Yoav Benjamini for his valuable contributions to the original research and for granting permission to use material from our technical paper.

\bibliography{main}

\appendix
\section{Toy Example Generation}
\label{app:toy_example_generation}

The data example is deliberately simple, using mainly Gaussian relationships and a variable that is not really participating in the system.

The example simulates the decision of how fast to turn a wind turbine, where the outcomes are energy produced and perceived noise. It is loosely based on research by \cite{Staffell2014}, who state:

\begin{quote}
``By accounting for individual site conditions we confirm that load factors do decline with age, at a similar rate to other rotating machinery. Wind turbines are found to lose 1.6 ± 0.2\% of their output per year, with average load factors declining from 28.5\% when new to 21\% at age 19. This trend is consistent for different generations of turbine design and individual wind farms. This level of degradation reduces a wind farm's output by 12\% over a twenty year lifetime, increasing the levelised cost of electricity by 9\%.''
\end{quote}

The variables in the toy example are generated as follows:

\begin{align*}
\text{Winter\_Ind} &\sim \text{Binomial}(n=1, p=0.7) \\
\text{Sea\_Temperature} &\sim \mathcal{N}(\mu = 20 - \text{Winter\_Ind} \cdot 10, \sigma = 2) \\
\text{Wind\_Speed} &\sim \mathcal{N}(\mu = 40 + \text{Winter\_Ind} \cdot 20, \sigma = 10) \\
\text{Strength\_Degradation} &\sim \mathcal{N}(\mu = 1.5, \sigma = 0.1) \\
\text{Rotational\_RPM} &\sim \mathcal{N}(\mu = 1.2 + \frac{\text{Wind\_Speed}}{10}, \sigma = 0.2) \\
\text{Energy\_Yield} &\sim \mathcal{N}(\mu = 10 + \frac{\text{Rotational\_RPM}}{1.5} - \frac{\text{Strength\_Degradation}}{10}, \sigma = 2) \\
\text{Perceived\_Noise} &\sim \mathcal{N}(\mu = 20 + \frac{\text{Rotational\_RPM}}{1.5} - \frac{\text{Wind\_Speed}}{4}, \sigma = 1)
\end{align*}

Where $\mathcal{N}(\mu, \sigma)$ denotes a normal distribution with mean $\mu$ and standard deviation $\sigma$.

Note:
\begin{itemize}
    \item Rotational\_RPM is measured in thousands (i.e., the actual RPM is 1000 times the value used in the model).
    \item Energy\_Yield is measured in Cent/kWh.
    \item Perceived\_Noise is an arbitrary scale.
\end{itemize}

The toy example was developed using Python and used capabilities provided by the DoWhy project [https://github.com/py-why/dowhy].

\end{document}